\documentclass
[prd,onecolumn,showpacs,floatfix,superscriptaddress,10pt]{revtex4-2}%
\usepackage{graphicx}
\usepackage{subfigure}
\usepackage{epsf}
\usepackage{bm}
\usepackage{amsmath}
\usepackage{amsfonts}
\usepackage{amssymb}
\usepackage{graphicx}
\usepackage{tabularx}
\usepackage{color}
\usepackage{tikz,xcolor,hyperref}%
\providecommand{\U}[1]{\protect\rule{.1in}{.1in}}

\newcommand{\be}{\begin{equation}}
\newcommand{\ee}{\end{equation}}

\newcommand{\mincir}{\raise
-3.truept\hbox{\rlap{\hbox{$\sim$}}\raise4.truept\hbox{$<$}\ }}
\newcommand{\magcir}{\raise
-3.truept\hbox{\rlap{\hbox{$\sim$}}\raise4.truept\hbox{$>$}\ }}

\providecommand{\U}[1]{\protect\rule{.1in}{.1in}}
\definecolor{lime}{HTML}{A6CE39}
\DeclareRobustCommand{\orcidicon}{	\begin{tikzpicture}
	\draw[lime, fill=lime] (0,0)
	circle [radius=0.16]
	node[white] {{\fontfamily{qag}\selectfont \tiny ID}};
	\draw[white, fill=white] (-0.0625,0.095)
	circle [radius=0.007];
	\end{tikzpicture}
	\hspace{-2mm}
}
\foreach \x in {A, ..., Z}{
	\expandafter\xdef\csname orcid\x\endcsname{\noexpand\href{https://orcid.org/\csname orcidauthor\x\endcsname}{\noexpand\orcidicon}}
}
\begin{document}
\title{Anisotropic Cosmology with Inhomogeneous $SU(2) $ and $SL(2,\mathbb{R})$
Nonlinear $\sigma$-Models}
\author{N. Dimakis}
\email{nikolaos.dimakis@ufrontera.cl}
\affiliation{Departamento de Ciencias F\'{\i}sicas, Universidad de La Frontera, Casilla
54-D, 4811186 Temuco, Chile}
\author{A. Giacomini}
\email{alexgiacomini@uach.cl}
\affiliation{Instituto de Ciencias Fisicas y Matem\'aticas, Universidad Austral de Chile,
Valdivia, Chile}
\author{G. Leon}
\email{genly.leon@ucn.cl}
\affiliation{Departamento de Matem\'{a}ticas, Universidad Cat\'{o}lica del Norte, Avda.
Angamos 0610, Casilla 1280 Antofagasta, Chile}
\affiliation{Institute of Systems Science, Durban University of Technology, PO Box 1334,
Durban 4000, South Africa}
\affiliation{Centre for Space Research, North-West University, Potchefstroom 2520, South
Africa. }
\author{F. Moncada}
\email{fmoncada@uct.cl}
\affiliation{Departamento de Ciencias Matem\'{a}ticas y F\'{\i}sicas, Facultad de
Ingenier\'{\i}a, Universidad Cat\'olica de Temuco, Temuco, Chile}
\author{A. Paliathanasis}
\email{anpaliat@phys.uoa.gr}
\affiliation{Institute of Systems Science, Durban University of Technology, Durban 4000,
South Africa}
\affiliation{Centre for Space Research, North-West University, Potchefstroom 2520, South Africa}
\affiliation{Centro de Investigaci\'on, Innovaci\'on y Creaci\'on (CIIC), Universidad
Cat\'olica de Temuco, Temuco, Chile}
\affiliation{Departamento de Ciencias Matem\'{a}ticas y F\'{\i}sicas, Facultad de
Ingenier\'{\i}a, Universidad Cat\'olica de Temuco, Temuco, Chile.}
\affiliation{National Institute for Theoretical and Computational Sciences (NITheCS), South Africa}

\begin{abstract}
We investigate the cosmological dynamics of the nonlinear $\sigma$-model in a
locally rotational symmetric Bianchi V background geometry. We employ the
$SU(2)$ and the $SL(2,\mathbb{R})$ internal symmetries, for which the kinetic
metric of the multi-scalar fields is a space of constant curvature. We assume
an inhomogeneous ansatz for the scalar fields, such that the $\sigma$-model,
which allows the existence of nontrivial solutions in the Bianchi V geometry.
For an exponential potential, we perform a detailed analysis of the phase
space, we determine the stationary points, and we study their stability
properties. The scalar field inhomogeneities prevent both the slow-roll and
hyperbolic inflationary epochs. For the $SL(2,\mathbb{R})$ internal symmetries
inflation is recovered only when the model is reduced to the single scalar
field theory. The existence of a cosmological constant is required to describe
cosmic acceleration in the case of $SU\left(  2\right)  $ symmetry.

\end{abstract}
\keywords{Cosmology; Scalar Field; Quintom; Dynamical analysis}
\pacs{98.80.-k, 95.35.+d, 95.36.+x}
\date{\today}
\maketitle

\section{Introduction}

Inflation is the main mechanism which explains long-standing problems on the
structure of the universe, such as the flatness and the horizon problems
without requiring fine tuned initial conditions \cite{Aref1,guth,ln1,rf1}.
During inflation, the universe is under a rapid accelerated phase, which
dynamically suppresses the effects of initial spatial curvature, anisotropies,
and inhomogeneities. Therefore, the influence of the initial conditions on the
cosmological evolution is neglected.

Scalar fields play a significant role for the description of inflation. For a
single scalar field minimally coupled to gravity, during the slow-roll epoch,
when the potential term dominates over the kinetic term, the degrees of
freedom of the scalar field can drive the gravitational dynamics to describe
cosmic acceleration \cite{Ratra}. A wide variety of inflationary potentials
has been proposed in the literature to account for the cosmological
observations \cite{ref1,ref2,ref3,newinf,ref4,ref5a,ln6,ln7,ln8}, while in
many cases Scalar fields are also applied as dark energy candidates to
describe the late-time acceleration \cite{Ratra,f0,f1,f2,f3,f4,f5,f6,f7}.

In more generic inflationary contexts, multi-scalar field models have been
introduced \cite{Brown:2017osf,Mizuno:2017idt}. They provide a richer
cosmological evolution than the standard single-field inflationary models. In
multi-scalar field inflation, the presence of additional scalar degrees of
freedom gives rise to entropy perturbations, which affect the curvature
perturbation. Consequently, the evolution of the curvature perturbation
depends on the number of e-folds. Furthermore, the evolutionary trajectory in
field space can be sensitive to the initial conditions, leading to different
physical conditions at the end of inflation \cite{mp1,mp2,mp3}. Finally,
multi-scalar field models provide a natural mechanism for the generation of
non-Gaussianities in the primordial power spectrum \cite{mp4,mp5}.

The nonlinear $\sigma$-model provides a mathematical framework for
constructing multi-scalar field theories, where the scalar fields evolve
dynamically in a nontrivial manifold. This manifold defines the structure of
the kinetic part for the scalar fields and its properties follow from the
algebraic structure employed to construct the model. A two-scalar field model
of special interest is the Chiral model \cite{ch1,ch2,ch3,ch4,ch5}, which
emerges from the use of the $SL\left(  2,\mathbb{R}\right)  $ Lie algebra
within the nonlinear $\sigma$-model framework. This yields a maximally
symmetric kinetic metric for the two scalar fields. The Chiral model possess
two inflationary epochs: the standard slow-roll inflation driven by a single
scalar field, and a new inflationary epoch where both scalar fields contribute
nontrivial in the cosmic fluid. The second phase is known as hyperbolic
inflation \cite{Brown:2017osf}.

The global dynamics of the field equations within an isotropic and homogeneous
Friedmann-Lema\^{\i}tre-Robertson-Walker (FLRW) cosmology for the Chiral
model, was investigated in \cite{ch3,ch6,ch7} for an exponential potential. In
\cite{anp1} the two-scalar field model was used to describe dark mater and
provide a mechanism for the unification of the dark sector. Exact analytical
solutions for the exponential potential and beyond were derived in
\cite{sol1,sol2}. In \cite{qun1}, it was demonstrated that the effective
equation of state parameter can cross the phantom divide line via a quantum
transition mechanism. Recently, the Chiral model \cite{chi1} was investigated
as a dark energy candidate and confronted with recent observational data. It
was found that it can challenge the standard $\Lambda$CDM cosmology, providing
a statistically fit comparable to the $w_{0}w_{a}$CDM model without requiring
a phantom crossing. The effects of anisotropies were investigated in
\cite{ans1}, where it was shown that the two inflationary epochs provided by
the two scalar fields sufficiently suppress cosmological anisotropies, so that
the late-time universe is well described by the spatially flat FLRW geometry,
in agreement with the cosmic \textquotedblleft no-hair\textquotedblright%
\ conjecture \cite{nh}.

Spatially homogeneous, but generally anisotropic, cosmological models are
classified into the nine Bianchi types, depending on the admitted
three-dimensional group of motions, $G_{3}$, acting simply transitively on the
spacelike hypersurfaces \cite{kramer}. The Bianchi models have been
extensively used to describe primordial anisotropies, the CMB polarization,
and other cosmological phenomena
\cite{jacobs2,collins,JB1,JB2,y1,y2,kas2,kas3,lf1,lf2,lf3}. The impact of
small anisotropies and nonzero spatial curvature at the early stages of the
universe on present-day cosmological observables has been investigated
previously in \cite{pl1,pl2,val1}. The Bianchi models are further classified
into two broad categories, namely Class A and Class B models. The
classification is based on the structure constants of the corresponding
three-dimensional Lie algebras. In terms of physics, the two families have
different geometrical and dynamical properties. In Class B models, there is a
coupling between the spatial curvature and the anisotropies, which leads to
additional constraints in the gravitational field equations. Moreover, the
energy-momentum tensor for the matter in Class B models can, in general,
include nonzero heat flux and anisotropic stress components. Thus, while a
perfect fluid matter source is naturally described in the comoving frame,
Class B models allow tilted fluid configurations, for which the fluid
four-velocity is not aligned with the comoving observer. Tilted cosmological
models within a Locally Rotationally Symmetric (LRS) Bianchi V geometry have
been studied in \cite{ke1,ke2,hew,col0,col0a,col0b}. In tilted cosmology, the
effective equation of state parameter for an ideal gas becomes time-varying
\cite{tt1}. Observational tests of tilted models using cosmological data have
also been presented in \cite{st1,st2}.

In this work, we consider the nonlinear $\sigma$-model with two internal
symmetry groups: $SU\left(  2\right)  $ and $SL\left(  2,\mathbb{R}\right)  $,
within a locally rotationally symmetric Bianchi V background geometry. This
framework allows the nonlinear $\sigma$-model to possess fewer symmetries than
the background geometry, thereby permitting the corresponding multi-scalar
field configuration to be inhomogeneous. The main objective of this study is
to investigate whether inflation can occur within the nonlinear $\sigma$-model
in a Bianchi V background, and to examine how scalar-field inhomogeneities
affect the inflationary dynamics. A class of inhomogeneous $\sigma$-models
with the $SU\left(  2\right)  $ internal symmetry was recently studied in
\cite{fc1,fc2,fc2b}, where it was found that it allows for the existence of
black holes with ``bumpy'' horizons. This framework was extended to
inhomogeneous cosmologies, where new classes of Szekeres-like exact solutions
were derived \cite{fc3}.

The structure of the papers follows: In Section \ref{sec2}, we introduce the
basic framework for the nonlinear $\sigma$-model and present the cosmological
model considered in this work, namely, a homogeneous and anisotropic Bianchi V
geometry. We introduce an inhomogeneous ansatz for the scalar fields, which
allows for nontrivial field configurations compatible with the background
geometry. Section \ref{sec3} contains the main analysis of this study, where
we investigate the phase-space structure of the cosmological field equations.
We derive the stationary points and analyze their stability properties. Based
on these results, we reconstruct the cosmological evolution associated with
the nonlinear $\sigma$-model. Finally, in Section \ref{sec4} we discuss the
obtained results and draw our conclusions.

\section{Inhomogeneous Nonlinear $\sigma$-model in anisotropic cosmology}

\label{sec2}

The nonlinear $\sigma$-model \cite{nc1,pol} arise at the leading order of
Chiral perturbation theory which provides the low-energy effective description
of QCD \cite{R11,pp1}. Within the $SU(2)$ consideration the Action Integral
for the nonlinear $\sigma$-model reads%
\begin{equation}
S_{NLSM}=\frac{1}{4}\int d^{4}x\sqrt{-g}tr\left(  U^{-1}\partial_{\mu
}U\right)  ^{2}\ . \label{ac.01}%
\end{equation}

\subsection{$SU\left(  2\right)  $ internal symmetry}

We assume the parametrization for the $SU\left(  2\right)  $ group to be
\cite{fc1}
\begin{equation}
U=\boldsymbol{1}_{2\times2}\cos\alpha+\sin\alpha\mathbf{n}_{j}\mathbf{\tau
}^{j}\ , \label{ac.02}%
\end{equation}
in which $\mathbf{n}$ denotes the isospin vector
\begin{equation}
\mathbf{n}=(\sin f\cos\gamma,\sin f\sin\gamma,\cos f)\ , \label{ac.03}%
\end{equation}
and $\tau^{j}=i \sigma^{j}$, where $\sigma^{j}$ are the Pauli matrices. The
$\alpha=\alpha\left(  x^{\mu}\right)  ,$ $f=f\left(  x^{\mu}\right)  $ and
$\gamma=\gamma\left(  x^{\mu}\right)  $ are the three scalar fields
parameterizing the group manifold.

Within this parametrization the action integral (\ref{ac.01}) is expressed as
\begin{equation}
S_{NLSM}=-\int d^{4}x\sqrt{-g}\left\{  \frac{1}{2}\left(  \nabla_{\mu}%
\alpha\right)  \left(  \nabla^{\mu}\alpha\right)  +\frac{1}{2}\sin^{2}%
\alpha\left[  \left(  \nabla_{\mu}f\right)  \left(  \nabla^{\mu}f\right)
+\sin^{2}f\left(  \nabla_{\mu}\gamma\right)  \left(  \nabla^{\mu}%
\gamma\right)  \right]  + V\left(  \alpha\right)  \right\}  \ , \label{ac.04}%
\end{equation}
where we have introduced the potential function $V\left(  \alpha\right)  $.
Variation with respect to the background metric leads to the energy-momentum
tensor $T_{\mu\nu}=\frac{-2}{\sqrt{-g}}\frac{\delta S_{NLSM}}{\delta g^{\mu
\nu}}$, which assumes the form \cite{fc1}%
\begin{align}
T_{\mu\nu}^{NLSM}  &  =\left(  \nabla_{\mu}\alpha\right)  \left(  \nabla_{\nu
}\alpha\right)  +\sin^{2}\alpha\left[  \left(  \nabla_{\mu}f\right)  \left(
\nabla_{\nu}f\right)  +\sin^{2}f\left(  \nabla_{\mu}\gamma\right)  \left(
\nabla_{\nu}\gamma\right)  \right] \nonumber\\
&  -g_{\mu\nu}\left[  \frac{1}{2}\left(  \nabla_{\kappa}\alpha\right)  \left(
\nabla^{\kappa}\alpha\right)  +\frac{1}{2}\sin^{2}\alpha\left[  \left(
\nabla_{\kappa}f\right)  \left(  \nabla^{\kappa}f\right)  +\sin^{2}f\left(
\nabla_{\kappa}\gamma\right)  \left(  \nabla^{\kappa}\gamma\right)  \right]
+V\left(  \alpha\right)  \right]  .
\end{align}
Variation of the action integral (\ref{ac.04}) with respect to the scalars
$\alpha$, $f$, and $\gamma$ respectively, leads to the equations of motion%
\begin{equation}
\square\Phi^{A}+\Gamma_{BC}^{A}\nabla_{\kappa}\Phi^{B}\nabla^{\kappa}\Phi
^{C}+V^{,A}\left(  \Phi^{C}\right)  =0,
\end{equation}
where $\square=\frac{1}{\sqrt{-g}}\frac{\partial}{\partial x^{\mu}}\left(
\sqrt{-g}g^{\mu\nu}\frac{\partial}{\partial x^{\nu}}\right)  $ is the
Laplacian for the background metric tensor $g_{\mu\nu},$ $\Phi^{A}=\left(
\alpha,f,\gamma\right)  ^{T}$, and $\Gamma_{BC}^{A}$ are the connection
coefficients of the metric tensor $\gamma_{AB}$, which defines the kinetic
term of the scalar-field Lagrangian (\ref{ac.04}), i.e. $\gamma_{AB}%
=\frac{\partial L_{NLSM}}{\partial\Phi^{A}\partial\Phi^{B}}$. The line element
of $\gamma_{AB}$ in the field space reads%
\begin{equation}
ds_{\gamma}^{2}=d\alpha^{2}+\sin^{2}\alpha\left[  df^{2}+\sin^{2}fd\gamma
^{2}\right]  ,
\end{equation}
and describes a space of constant positive curvature.

The equations for the three scalar fields are \cite{fc3}
\begin{equation}
-\square\alpha+\frac{\sin(2\alpha)}{2}\left[  \left(  \nabla_{\kappa}f\right)
\left(  \nabla^{\kappa}f\right)  +\sin^{2}f\left(  \nabla_{\kappa}%
\gamma\right)  \left(  \nabla^{\kappa}\gamma\right)  \right]  +V_{,\alpha
}=0\ , \label{equ1}%
\end{equation}%
\begin{equation}
-\square f-2\cot\alpha\left(  \nabla_{\kappa}\alpha\nabla^{\kappa}f\right)
+\frac{1}{2}\sin(2f)\left(  \nabla_{\kappa}\gamma\right)  \left(
\nabla^{\kappa}\gamma\right)  =0\ ,
\end{equation}%
\begin{equation}
-\square\gamma-2\cot\alpha\left(  \nabla_{\kappa}\alpha\nabla^{\kappa}%
\gamma\right)  -2\cot f\left(  \nabla_{\kappa}f\nabla^{\kappa}\gamma\right)
=0. \label{equ3}%
\end{equation}

\subsection{$SL\left(  2,\mathbb{R}\right)  $ internal symmetry}

For the internal symmetry given by the $SL\left(  2,\mathbb{R}\right)  $
group, the nonlinear $\sigma$-model reduces to the Chiral two-scalar field
model \cite{ch1}, characterized by $\alpha$ and $\gamma$. The field $U$ in
this case is
\begin{equation}
U = \left(
\begin{array}
[c]{cc}%
e^{\alpha} & e^{\alpha} \gamma\\
e^{\alpha} \gamma & e^{\alpha} \gamma^{2}+e^{-\alpha}\\
&
\end{array}
\right)  ,
\end{equation}
and the resulting action for the model reads (\eqref{ac.01} action should be
used with a minus in front in this case)
\begin{equation}
S_{NLSM}=-\int d^{4}x\sqrt{-g}\left\{  \frac{1}{2}\left(  \nabla_{\mu}%
\alpha\right)  \left(  \nabla^{\mu}\alpha\right)  +\frac{1}{2}e^{2\alpha
}\left(  \nabla_{\mu}\gamma\right)  \left(  \nabla^{\mu}\gamma\right)  +
V\left(  \alpha\right)  \right\}  . \label{ac.04}%
\end{equation}
The corresponding energy-momentum tensor is given by
\begin{align}
T_{\mu\nu}^{NLSM}  &  =\left(  \nabla_{\mu}\alpha\right)  \left(  \nabla_{\nu
}\alpha\right)  +e^{2\alpha}\left(  \nabla_{\mu}\gamma\right)  \left(
\nabla_{\nu}\gamma\right) \nonumber\\
&  -g_{\mu\nu}\left[  \frac{1}{2}\left(  \nabla_{\kappa}\alpha\right)  \left(
\nabla^{\kappa}\alpha\right)  +\frac{1}{2}e^{2\alpha} \left(  \nabla_{\kappa
}\gamma\right)  \left(  \nabla^{\kappa}\gamma\right)  +V\left(  \alpha\right)
\right]  .
\end{align}

The resulting equations of motion for the two scalar fields are
\begin{equation}
-\square\alpha+e^{2\alpha}\left(  \nabla_{\kappa}\gamma\right)  \left(
\nabla^{\kappa}\gamma\right)  +V_{,\alpha}=0,
\end{equation}%
\begin{equation}
-\square\gamma-2\left(  \nabla_{\kappa}\alpha\nabla^{\kappa}\gamma\right)  =0.
\end{equation}
In this model, the kinetic metric for the scalar fields is defined by the line
element
\begin{equation}
ds^{2}=d\alpha^{2}+e^{2\alpha}d\gamma^{2},
\end{equation}
which describes a space constant negative curvature.

\subsection{Bianchi\ V Background}

For the base manifold geometry $g_{\mu\nu}$ we consider the
homogeneous and anisotropic locally rotationally symmetric spacetime
\cite{ke1,ke2} 
\begin{equation}
ds^{2}=-dt^{2}+a\left(  t\right)  ^{2}\left(  e^{4b\left(  t\right)  }%
dx_{1}^{2}+e^{-2b\left(  t\right)  }e^{-2x_{1}}\left(  dx_{2}^{2}+dx_{3}%
^{2}\right)  \right)  . \label{tl.01}%
\end{equation}
where now $x^{\mu}=\left(  t,x,y,z\right)  ^{T}\,$. This is a
Bianchi type V metric. For arbitrary scale factors $a\left(  t\right)
$ and $b\left(  t\right)  $, it admits a four-dimensional
group of isometries generated by the Killing vectors 
\begin{align*}
\xi_{1}  &  =\partial_{x_{1}}+x_{2}\partial_{x_{2}}+x_{3}\partial_{x_{3}}~,\\
\xi_{2}  &  =\partial_{x_{2}},\\
\xi_{3}  &  =\partial_{x_{3}},\\
\xi_{4}  &  =x_{3}\partial_{x_{2}}-x_{2}\partial_{x_{3}}%
\end{align*}
satisfying the Lie algebra, with nonzero commutators
\begin{align*}
\left[  \xi_{1},\xi_{3}\right]   &  =-\xi_{1}~,~\left[  \xi_{2},\xi
_{3}\right]  =-\xi_{2}~,~\\
\left[  \xi_{2},\xi_{4}\right]   &  =-\xi_{3},~\left[  \xi_{3},\xi_{4}\right]
=\xi_{2}%
\end{align*}
where the the three-dimensional subalgebra $\left\{  \xi_{1},\xi
_{2},\xi_{3}\right\}$ belongs to the Class B Bianchi
classification.

Within the context of General Relativity as a gravitational theory, we write
the action integral
\begin{equation}
S=S_{EH}+S_{NLSM}+S_{m},
\end{equation}
where $S_{EH}$ is the Einstein-Hilbert action, $S_{EH}=\int d^{4}x\sqrt
{-g}\frac{R}{2}$, with $S_{m}$ describing a dust fluid component. The
equations describing the dynamics for the scale factors $a\left(  t\right)  $
and $b\left(  t\right)  $, which also involve the scalar fields $\alpha\left(
x^{\mu}\right)  ,~f\left(  x^{\mu}\right)  ,~\gamma\left(  x^{\mu}\right)  $
and the energy density for the dust fluid, are obtained from Einstein's
equations:
\begin{equation}
G_{\mu\nu}=T_{\mu\nu}^{NLSM}+T_{\mu\nu}^{m},
\end{equation}
where $T_{\mu\nu}^{m}=\frac{-2}{\sqrt{-g}}\frac{\delta S_{m}}{\delta g^{\mu
\nu}}$.

For the Bianchi V geometry (\ref{tl.01}) the nonzero components of the mixed
Einstein tensor $G^{\mu}_{\nu}$ are \cite{ke1,ke2}%

\begin{align*}
G_{\;t}^{t}  &  =3\left(  \dot{b}^{2}-H^{2}+\frac{e^{-4b}}{a^{2}}\right)  ,\\
G_{\;x}^{t}  &  =-a^{2} e^{4b} G_{\;t}^{x} =6\dot{b},\\
G_{\;x}^{x}  &  =-3\left(  H-\dot{b}\right)  ^{2}-2\dot{H}+2\ddot{b}%
+\frac{e^{-4b}}{a^{2}},\\
G_{\;y}^{y}  &  =G_{\;z}^{z} =-3\left(  H^{2}+H\dot{b}+\dot{b}^{2}\right)
-2\dot{H}-\ddot{b}+\frac{e^{-4b}}{a^{2}},
\end{align*}
where we define the Hubble function$H=\dot{a}/a$, with the dot denoting the
time derivative, i.e. $\dot{a}=\frac{da}{dt}$.

For the matter source, we assume that it is homogeneous, and defined within
the comoving frame, that is, the energy momentum tensor $T_{\mu\nu}^{m}$ has
the nonzero component
\begin{equation}
T_{tt}^{m}=\rho_{m}.
\end{equation}
It remains to determine the corresponding components of the nonlinear $\sigma
$-model's energy-momentum tensor, which we analyze in the following section.

\subsection{Inhomogeneous Nonlinear $\sigma$-Model}

For the nonlinear $\sigma-$model, let us consider the inhomogeneous ansatz%
\begin{equation}
\alpha=\alpha(t)\ ,\ \ f=\frac{\pi}{2}\ ,\ \ \gamma(t,x,y,z)=\gamma\left(
t\right)  +\sqrt{6}\gamma_{0}x,
\end{equation}
where $\gamma_{0}$ is a real number. Note that with a slight abuse of
notation, we use the same symbol $\gamma(t)$ for the temporal part of the
scalar field $\gamma(t,x,y,z)$; hereafter, $\gamma$ refers exclusively to
$\gamma(t)$. Parameter $\gamma_{0}$ is an inhomogeneous
parameter.

The resulting energy momentum tensor $T_{\mu\nu}^{NLSM}$ has the following
nonzero components%
\begin{align}
T_{tt}^{NLSM}  &  =\frac{1}{2}\left(  \dot{\alpha}^{2} +M^{2}\left(
\alpha\right)  ~\dot{\gamma}^{2}\right)  + V\left(  \alpha\right)  +
\frac{3e^{-4b}}{a^{2}}\gamma_{0}^{2}M^{2}\left(  \alpha\right)  ,\\
T_{tx}^{NLSM}  &  =\sqrt{6}\gamma_{0}M^{2} \left(  \alpha\right)  ~\dot
{\gamma},\\
T_{xx}^{NLSM}  &  =\frac{e^{4b} a^{2}}{2}\left(  \dot{\alpha}^{2}+M^{2}\left(
\alpha\right)  ~\dot{\gamma}^{2}\right)  - e^{4b} a^{2}V\left(  \alpha\right)
+3\gamma_{0}^{2}M^{2}\left(  \alpha\right)  ,\\
T_{zz}^{NLSM}  &  =T_{yy}^{NLSM}= -e^{-2(x+3 b)}T_{xx}- e^{-2(x+b)}a^{2}
\left(  2V-\dot{\alpha}^{2}- M^{2}(\alpha)\dot{\gamma}^{2}\right)  ,
\end{align}
in which $M\left(  \alpha\right)  =\sin\alpha$ or $M\left(  \alpha\right)
=e^{\alpha}$ for the $SU\left(  2\right)  $ and $SL\left(  2,\mathbb{R}%
\right)  $ internal symmetries respectively.

The Einstein field equations are now reduced to
\begin{align}
\dot{H}  &  =-\frac{3}{2}\left(  H^{2}+\dot{b}^{2}\right)  -\frac{1}{2}\left(
\frac{1}{2}\left(  \dot{\alpha}^{2} +M^{2}\left(  \alpha\right)  ~\dot{\gamma
}^{2}\right)  -V\left(  \alpha\right)  -\frac{e^{-4b}}{a^{2}}\left(
1+\gamma_{0}^{2}M^{2}\left(  \alpha\right)  \right)  \right)  ,\label{fe.01}\\
\ddot{b}  &  =-3H\dot{b}+2\gamma_{0}^{2}M^{2}\left(  \alpha\right)
\frac{e^{-4b}}{a^{2}}, \label{fe.02}%
\end{align}
with the constraint equations
\begin{equation}
3H^{2}=3\dot{b}^{2}+\frac{1}{2}\left(  \dot{\alpha}+M^{2}\left(
\alpha\right)  ~\dot{\gamma}^{2}\right)  +V\left(  \alpha\right)
+\frac{3e^{-4b}}{a^{2}}\left(  1+\gamma_{0}^{2}M^{2}\left(  \alpha\right)
\right)  +\rho_{m}, \label{fe.03}%
\end{equation}
and%
\begin{equation}
\frac{e^{-2b}}{a^{2}}\left(  \sqrt{6}\dot{b}+\gamma_{0}M^{2}\left(
\alpha\right)  \dot{\gamma}\right)  =0 , \label{fe.04}%
\end{equation}
which represent the quadratic and the linear constraint respectively. The
equations of motion for the remaining scalar fields $\alpha\left(  t\right)  $
and $\gamma\left(  t\right)  $ are
\begin{align}
\ddot{\alpha}  &  =-3H\dot{\alpha}+\frac{1}{2}\sin\left(  2\alpha\right)
\dot{\gamma}^{2}-V_{,\alpha}-3\gamma_{0}^{2}\sin\left(  2\alpha\right)
\frac{e^{-4b}}{a^{2}},\label{fe.05}\\
\ddot{\gamma}  &  =-3H\dot{\gamma}-2\cot\alpha~\dot{\alpha}\dot{\gamma}%
-2\sqrt{6}\gamma_{0}\frac{e^{-4b}}{a^{2}}. \label{fe.06}%
\end{align}
for the $SU\left(  2\right)  $ symmetry, and
\begin{align}
\ddot{\alpha}  &  =-3H\dot{\alpha}+e^{2\alpha}\dot{\gamma}^{2}-V_{,\alpha
}-6\gamma_{0}^{2}e^{2\alpha}\frac{e^{-4b}}{a^{2}},\\
\ddot{\gamma}  &  =-3H\dot{\gamma}-2\dot{\alpha}\dot{\gamma}-2\sqrt{6}%
\gamma_{0}\frac{e^{-4b}}{a^{2}}.
\end{align}
for the $SL\left(  2,\mathbb{R}\right)  $ symmetry. Finally, the equation of
motion for the matter source is given by
\begin{equation}
\dot{\rho}_{m}=-3H\rho_{m}\text{.}%
\end{equation}

We remark that a nonzero value $\gamma_{0}$, is a requirement in order for
anisotropic solutions to exist. For a homogeneous nonlinear $\sigma$-model,
where $\gamma_{0}=0$, it follows from (\ref{fe.04}) that $b=b_{0}$ or
$\frac{e^{-2b}}{a^{2}}\rightarrow0$, which implies that the resulting line
element (\ref{tl.01}) is that of the open FLRW geometry or the Bianchi I at
the asymptotic limit, respectively. Let us note that a nonzero $\gamma_{0}$
\ introduces a coupling between the scalar field $\alpha$ and the spatial
curvature for the background metric. Indeed, from Eq. (\ref{fe.05}) we can
read the effective mass term
\begin{equation}
m_{\alpha,eff}^{2}=V_{,\alpha\alpha}+\cos\left(  2\alpha\right)  \left(
6\gamma_{0}^{2}\frac{e^{-4b}}{a^{2}}-\dot{\gamma}^{2}\right)  ,
\end{equation}
or%
\begin{equation}
m_{\alpha,eff}^{2}=V_{,\alpha\alpha}+2e^{2\alpha}\left(  6\gamma_{0}^{2}%
\frac{e^{-4b}}{a^{2}}-\dot{\gamma}^{2}\right)  .
\end{equation}

In what follows, we investigate the cosmological dynamics for the nonlinear
$\sigma$-model and, in particular, the role of inhomogeneities in the
inflationary dynamics. To this end, we consider an exponential potential,
$V\left(  \alpha\right)  =V_{0}e^{\lambda\alpha}$, which provides a simple and
well-motivated parametrization of the scalar-field potential and has been
extensively studied in the context of inflationary and cosmological dynamics.
The exponential form is a particularly useful choice in the dynamical system
analysis of gravitational physics, as it captures the asymptotic behavior of a
more general hyperbolic potential.

\section{Asymptotic Dynamics}

\label{sec3}

To investigate the cosmological dynamics of the nonlinear $\sigma$-model and
to identify its asymptotic solutions, we recast the field equations into the
equivalent form of an autonomous system of dimensionless variables. We employ
the Hubble normalization approach \cite{cop1,cop2} and we introduce the
following dependent variables for the scalar field%
\[
x=\frac{\dot{\alpha}}{\sqrt{6}H},~y=\sqrt{\frac{V\left(  \alpha\right)
}{3H^{2}}},~z=\frac{M\left(  \alpha\right)  \dot{\gamma}}{\sqrt{6}H},~
\]
together with
\[
\Omega_{m}=\frac{\rho_{m}}{3H^{2}},~\Sigma=\frac{\dot{b}}{H}~,~W=\frac
{e^{-2b}}{aH}\sqrt{1+\gamma_{0}^{2}M^{2}\left(  \alpha\right)  },
\]
for the background component. Parameter $W$ corresponds to the effective
spatial curvature term, where now, due to the inhomogeneous scalar field there
is a nonlinear coupling between the spatial curvature and the scalar field
$\alpha$. Furthermore, under the independent variable $\tau=\ln a$, the field
equations are rewritten as follows
\begin{equation}
\frac{dx}{d\tau}=\frac{1}{2}\left(  \sqrt{6}\left(  2\frac{M_{,\alpha}} {M}
z^{2}-\lambda y^{2}-\frac{2\gamma_{0}^{2}W^{2}M_{,\alpha}M}{1+\gamma_{0}%
^{2}M^{2}}\right)  +x\left(  3\left(  x^{2}-1-y^{2}+z^{2}+\Sigma^{2}\right)
-W^{2}\right)  \right)  , \label{ff.01}%
\end{equation}

\begin{equation}
\frac{dy}{d\tau}=\frac{1}{2}y\left(  \sqrt{6}\lambda x+3\left(  1+x^{2}%
-y^{2}+z^{2}+\Sigma^{2}\right)  -W^{2}\right)  ,\label{ff.02}%
\end{equation}%
\begin{equation}
\frac{dz}{d\tau}=\frac{1}{2}\left(  z\left(  3\left(  x^{2}-1-y^{2}%
+z^{2}+\Sigma^{2}\right)  -2\sqrt{6}x\frac{M_{,\alpha}}{M}\right)
-W^{2}\left(  z+\frac{4\gamma_{0}M}{1+\gamma_{0}^{2}M^{2}}\right)  \right)
,\label{ff.03}%
\end{equation}%
\begin{equation}
\frac{d\alpha}{d\tau}=\sqrt{6}x,\label{ff.04}%
\end{equation}
and%
\begin{equation}
\frac{d\Sigma}{d\tau}=\frac{3}{2}\Sigma\left(  x^{2}-1-y^{2}+z^{2}+\Sigma
^{2}\right)  -\frac{1}{2}\Sigma W^{2}+\frac{2\gamma_{0}^{2}W^{2}M^{2}%
}{1+\gamma_{0}^{2}M^{2}},\label{ff.05}%
\end{equation}%
\begin{equation}
\frac{dW}{d\tau}=-\frac{1}{2}W\left(  W^{2}-1-3\left(  x^{2}-y^{2}%
+z^{2}\right)  +\left(  4-3\Sigma\right)  \Sigma\right)  +\frac{\sqrt{6}}%
{2}\frac{2\gamma_{0}^{2}xWM_{,\alpha}M}{1+\gamma_{0}^{2}M\left(
\alpha\right)  ^{2}}.\label{ff.06}%
\end{equation}

Furthermore, the constraint equations (\ref{fe.03}), (\ref{fe.04}) lead to the
algebraic equations%
\begin{equation}
W\left(  \Sigma+\gamma_{0}zM\right)  =0,\label{ff.07}%
\end{equation}
and
\begin{equation}
1-x^{2}-y^{2}-z^{2}-\Sigma^{2}-W^{2}=\Omega_{m}.\label{ff.08}%
\end{equation}

Furthermore, for the effective mass term, we define the dimensionless
variable $\bar{m}_{\alpha,eff}^{2}=\frac{m_{\alpha,eff}^{2}}{3H^{2}}%
$, as follows
\begin{equation}
\bar{m}_{\alpha,eff}^{2}=\lambda^{2}y^{2}+3\left(  M^{2}\right)  _{,a\alpha
}\left(  \frac{\gamma_{0}}{1+\gamma_{0}^{2}M^{2}}-\frac{z^{2}}{M^{2}}\right)
\text{.} \label{ff.10}%
\end{equation}

The asymptotic cosmological solutions can be identified with the critical
points of the above dynamical system and their stability can be investigated
by analyzing the corresponding linearized equations. From (\ref{ff.08})
together with the physical requirement for the dust-fluid component,
$0\leq\Omega_{m}\leq1$, it follows that the normalized variables are subject
to certain bounds; that is, $\left\{  x,y,z,\Sigma,W\right\}  $ take values
within the unit ball in $\mathbb{R}^{5}$, while variable $\alpha$ is
constrained as $\alpha\in(-\frac{\pi}{2},\frac{\pi}{2}]$ for the $SU\left(
2\right)  $ case. We further observe that the dynamical system possesses
discrete symmetries under the transformations, $y\rightarrow-y$, or
$W\rightarrow-W$. Consequently, the dynamics in the corresponding negative
sectors is equivalent to that in the positive sectors. Without loss of
generality, we may therefore restrict our analysis to positive $\left\{
y,W\right\}  $ variables. Hence, the relevant phase space is a bounded subset
of $\mathbb{R}^{5}$ as follows
\begin{align*}
-1  &  \leq x\leq1~,~0\leq y\leq1~,~-1\leq z\leq1\\
-1  &  \leq\Sigma\leq1~,~0\leq W\leq1\text{.}%
\end{align*}
Finally, it is important to clarify that due to the algebraic constraint
equation (\ref{ff.07}) the dimension of the dynamical system is reduced by
one. In the absence of a matter term, Eq. (\ref{ff.08}) reduces further the
dimensionality of the dynamical system.

Each stationary point of the dynamical system describes an asymptotic
solution, with value of the deceleration parameter $q=-1-\frac{\dot{H}}{H^{2}%
}$ as derived by the expression
\begin{equation}
q\left(  x,y,z,\Sigma,w,\alpha\right)  =\frac{1}{2}\left(  1+3\left(
x^{2}-y^{2}+z^{2}+\Sigma^{2}\right)  -W^{2}\right)  .
\end{equation}
Let us proceed with the detailed analysis of each $\sigma$-model.

\subsection{$SU\left(  2\right)  $ symmetry}

Consider the case with the $SU\left(  2\right)  $ internal symmetry. The
stationary points $P=P\left(  x,y,z,\Sigma,W,\alpha\right)  $ are summarized
in four families.

The first family consists of the one-parameter stationary point%
\[
P_{1}=\left(  0,0,0,0,0,\alpha\right)  ,
\]
where $\alpha$ is arbitrary. At this point, the matter density parameter and
the deceleration parameter are $\Omega_{m}\left(  P_{1}\right)  =1$, and
$q\left(  P_{1}\right)  =\frac{1}{2}$. Therefore, $P_{1}$ corresponds to the
standard isotropic matter dominated FLRW solution.

The second family is composed of four stationary points,
\begin{align*}
P_{2}  &  =\left(  0,0,\sqrt{\Sigma\left(  1-\Sigma\right)  },\Sigma
,\sqrt{1-\Sigma},\arctan\sqrt{\frac{\Sigma}{\gamma_{0}^{2}\left(
1-\Sigma\right)  -\Sigma}}\right)  ,\\
P_{3}  &  =\left(  0,0,-\sqrt{\Sigma\left(  1-\Sigma\right)  },\Sigma
,\sqrt{1-\Sigma},\arctan\sqrt{\frac{\Sigma}{\gamma_{0}^{2}\left(
1-\Sigma\right)  -\Sigma}}\right)  ,\\
P_{4}  &  =\left(  0,0,\sqrt{1-\Sigma^{2}},\Sigma,0,\frac{\pi}{2}\right)  ,\\
P_{5}  &  =\left(  0,0,-\sqrt{1-\Sigma^{2}},\Sigma,0,\frac{\pi}{2}\right)  .
\end{align*}
For all four points $P_{I}$ with $I=2,\ldots,5$, we have $\Omega_{m}\left(
P_{I}\right)  =0$, so these solutions correspond to matter-free regimes. For
these points, the parameter $\Sigma$ remains arbitrary, and it is constrained
only from the existence conditions of the corresponding stationary point.
Specifically, $P_{2}$, and $P_{3}$\ are real and physically accepted when
$0\leq\Sigma\leq\frac{\gamma_{0}^{2}}{1+\gamma_{0}^{2}}$. Their deceleration
parameters are $q\left(  P_{2}\right)  =q\left(  P_{3}\right)  =2\Sigma$ and
$q\left(  P_{4}\right)  =q\left(  P_{5}\right)  =2$. Consequently, none of
these stationary points can describe an accelerated cosmological regime. A
further distinction between these solutions is provided by the anisotropy and
curvature variables. For $P_{2}$ and $P_{3}$, it holds $\Sigma W\neq0$
suggesting that both the anisotropic shear and the spatial curvature
contribution are non vanishing. Hence, the asymptotic solution at these points
describes anisotropic Bianchi V geometries. In contrast, for $P_{4}$ and
$P_{5}$ we have $W=0$, which implies that the points describe Bianchi I geometries.

We remark that while the contribution of the potential component to
the cosmological fluid is zero, the effective mass term $\bar{m}%
_{\alpha,eff}^{2}~$ is nonzero.

Moreover, we find the stationary points
\[
P_{6}=\left(  0,0,-\frac{1}{4\gamma_{0}},\frac{1}{4},\frac{\sqrt{3\left(
1+\gamma_{0}^{2}\right)  }}{4\gamma_{0}},\frac{\pi}{2}\right)  ,
\]
which describes a Bianchi V geometry where the nonlinear $\sigma$-model track
the dust fluid source, that is, with $\Omega_{m}\left(  P_{6}\right)
=\frac{1}{4}\left(  3-\frac{1}{\gamma_{0}^{2}}\right)  $ and $q\left(
P_{6}\right)  =\frac{1}{2}$.

The fourth family consists of the stationary point $P_{7}$ with coordinates%
\[
P_{7}=\left(  0,\frac{\sqrt{1+\lambda^{2}+3\gamma_{0}^{2}}}{\sqrt{2}\gamma
_{0}},\frac{\sqrt{1+\lambda^{2}}}{2\gamma_{0}},-\frac{1}{2},\frac{1}{2}%
\sqrt{-3\left(  1+\frac{1+\lambda^{2}}{\gamma_{0}^{2}}\right)  }%
,\operatorname{arccot}\lambda\right)  .
\]
However $P_{7}$ is always complex.\ Therefore, $P_{7}$ does not belong to the
real phase space and is consequently discarded as a physically acceptable
cosmological solution.

A special case follows when $\lambda=0$, for which the potential $V\left(
\alpha\right)  $ reduces to the limit of the the cosmological constant. An
additional stationary point appears%
\[
P_{8}=\left(  0,1,0,0,0,\alpha\right)  ,
\]
with $\alpha$ arbitrary, and physical quantities$~q\left(  P_{8}\right)  =-1$
and $\Omega_{m}\left(  P_{8}\right)  =0$. The asymptotic solution at $P_{8}%
$~describes the de Sitter accelerated universe.

\subsubsection{Stability properties}

To investigate the stability of the stationary points, we first use the
constraint equation (\ref{ff.07}) to eliminate the variable $\Sigma$, as in
general we assume that $W$ is not identically zero. The resulting system is
therefore reduced to a five-dimensional autonomous dynamical system with
dynamical variables $\left\{  x,y,z,W,\alpha\right\}  $. The Jacobian matrix
of the reduced dynamical system is then calculated at each stationary point.
Then from the eigenvalues of the Jacobian matrix, at each point, we can infer
the corresponding stability properties. A stationary point is characterized as
an attractor and the solution is stable if all the eigenvalues have negative
real parts. For positive real parts of the corresponding eigenvalues, the
point is characterized as a source, while for mixed eigenvalues the point is a
characterized as a saddle.

For $P_{1}$ we derive the eigenvalues $\left\{  -\frac{3}{2},-\frac{3}%
{2},\frac{3}{2},\frac{1}{2},0\right\}  $. From this we conclude that $P_{1}$
is always a saddle point. For points $P_{2}$ and $P_{3}$ one of the
eigenvalues is $1+2\Sigma$, where $\Sigma$ is a constant within the range
$0\leq\Sigma\leq\frac{\gamma_{0}^{2}}{1+\gamma_{0}^{2}}$, thus the Bianchi V
asymptotic solutions are always unstable. For the rest point of the second
family of asymptotic solutions, that is, $P_{4}$ and $P_{5}$ we find the
eigenvalues
\[
\frac{3}{2}\left(  1+\left(  1+\gamma_{0}^{2}\right)  \left(  1-\Sigma
^{2}\right)  \right)  ,~\frac{1}{2}\left(  -3+9\left(  1+\gamma_{0}%
^{2}\right)  \Sigma^{2}\right)  ~,~\frac{1}{2}\left(  1+\left(  1+\gamma
_{0}^{2}\right)  \Sigma^{2}+4\gamma_{0}\sqrt{1-\Sigma^{2}}\right)  ~~
\]%
\[
\frac{1}{4}\left(  3\gamma_{0}^{2}-3\left(  1+\gamma_{0}^{2}\right)
\Sigma^{2}\pm\sqrt{9\left(  \Sigma^{2}-\gamma_{0}^{2}\left(  1-\Sigma
^{2}\right)  \right)  ^{2}-96\left(  1-\Sigma^{2}\right)  }\right)  .
\]
Two of the latter eigenvalues are always positive, thus the points are
always unstable describing sources or saddle points. We find that the points
are sources when $\frac{1}{3\left(  1+\gamma_{0}^{2}\right)  }<\Sigma
^{2}<\frac{\gamma_{0}^{2}}{1+\gamma_{0}^{2}}$, with $\gamma_{0}%
>\frac{1}{3}$, otherwise, the points are saddle points. 

Moreover, for the matter solution $P_{6}$, the eigenvalues are
\[
\frac{3}{2},~\frac{\sqrt{3}}{4}\left(  -\sqrt{3}\pm\frac{\sqrt{9\gamma_{0}%
^{4}+7\gamma_{0}^{2}-2}}{\gamma_{0}\sqrt{1+\gamma_{0}^{2}}}\right)
,~\frac{\sqrt{3}}{4}\left(  -\sqrt{3}\pm\frac{\sqrt{2-\gamma_{0}^{2}%
-3\gamma_{0}^{4}}}{\gamma_{0}\sqrt{1+\gamma_{0}^{2}}}\right)  ,
\]
from where we conclude that the Bianchi V solution is always unstable.

Finally, for the de Sitter solution described by point $P_{8}$ it follows
$\left\{  -3,-3,-3,-1,0\right\}  $, where the zero eigenvalue is on the
direction which drives the dynamics of parameter $\alpha$. The latter does not
affect the cosmological model and the dynamical system can be separated.
Hence, the de Sitter solution is a future attractor for the dynamical system.

\begin{figure}[ptbh]
\centering\includegraphics[width=1\textwidth]{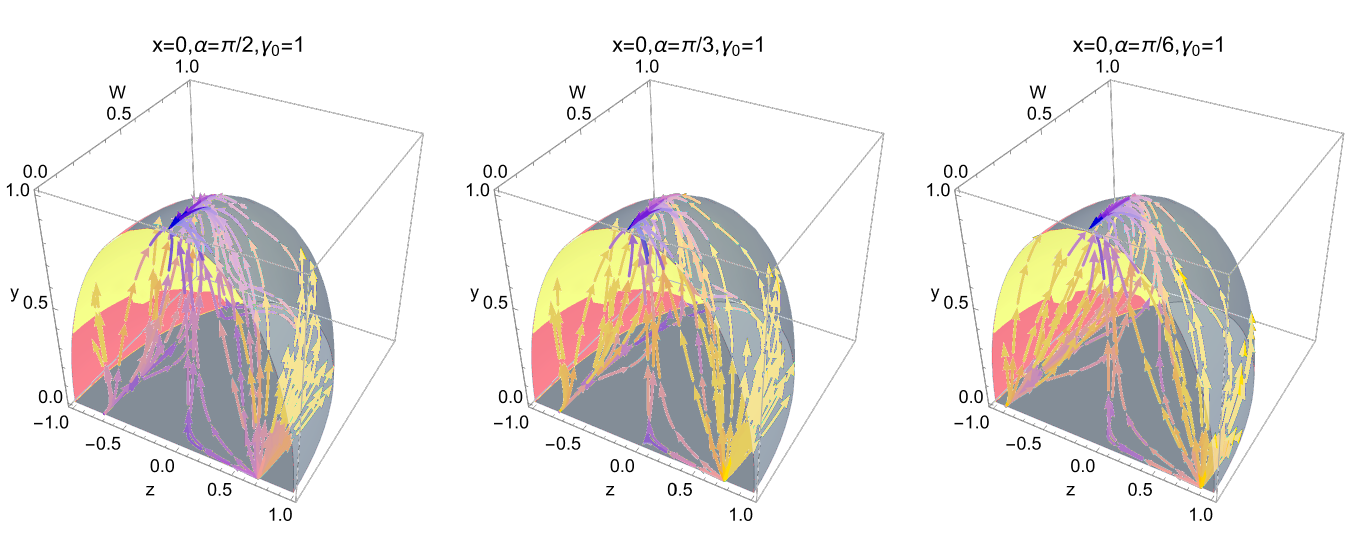}\caption{Phase-space
diagram for the nonlinear $\sigma$-model within the three-dimensional space
$\left\{  y,z,W\right\}  $ on the constant surfaces $x=0$ and $\alpha=\left(
\frac{\pi}{2},\frac{\pi}{4},\frac{\pi}{6}\right)  $. We observe that the
unique attractor is the point $\left(  1,0,0\right)  $, which is the de Sitter
point of the system. On the constant surface $x=0$, it is always a stationary
point without necessarily imposing $\lambda=0$. }%
\label{fig1}%
\end{figure}

In Fig. \ref{fig1} we present phase-space diagrams for the dynamical system
within the surface $x=0$, and $\alpha=\frac{\pi}{2}$.~Point $\left(
1,0,0\right)  $, which describes a de Sitter solution, is the unique attractor
in this three-dimensional submanifold. Recall that on the constant surface
$x=0$, the system is independent on parameter $\lambda$.

\begin{figure}[ptbh]
\centering\includegraphics[width=1\textwidth]{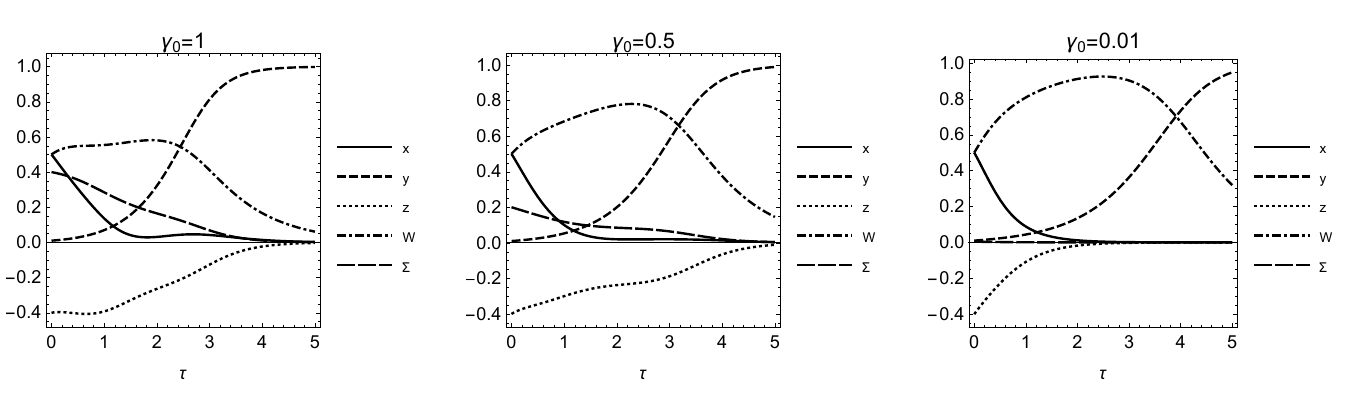}\caption{Dynamical
evolution of the dynamical variables$~\left\{  x,y,z,\Sigma,W\right\}  $ for
different values of parameter $\gamma_{0}$ and $\lambda=0$. }%
\label{fig2}%
\end{figure}

\begin{figure}[ptbh]
\centering\includegraphics[width=0.75\textwidth]{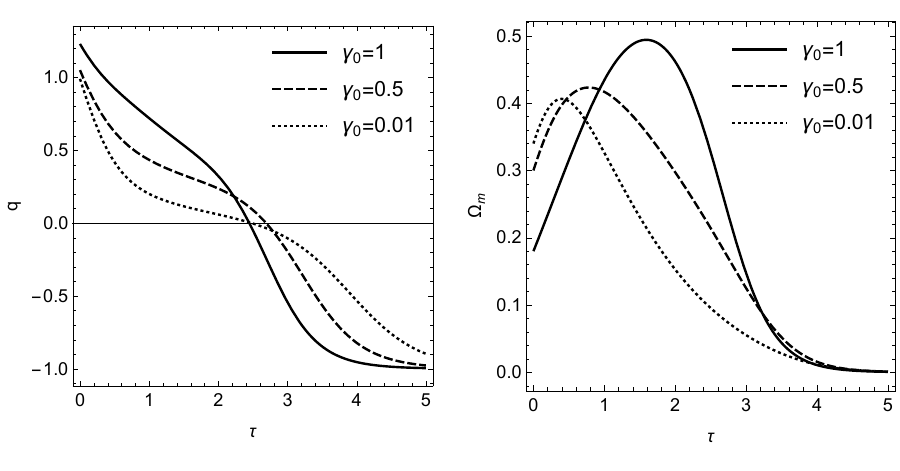}\caption{Qualitative
evolution of the physical variables $q$ and $\Omega_{m}$ for the numerical
solutions presented in Fig. \ref{fig2}.}%
\label{fig3}%
\end{figure}

For the case of the cosmological constant, that is, $\lambda=0$, in Fig.
\ref{fig2} we present the qualitative evolution of the dynamical variables
$\left\{  x,y,z,\Sigma,W\right\}  $ for different values of parameter
$\gamma_{0}$ which measure the inhomogeneity. The unique attractor is the de
Sitter point $P_{7}$ as can be verified and from the qualitative evolution of
the physical parameters $q$ and $\Omega_{m}$ as presented in Fig. \ref{fig3}.

\subsection{$SL\left(  2,\mathbb{R}\right)  $ symmetry}

For the Chiral model, the stationary points $Q=Q\left(  x,y,z,\Sigma
,W,\alpha\right)  $ are
\begin{align*}
Q_{1}  &  =\left(  0,0,0,0,0,\alpha\right)  ,\\
Q_{2}  &  =\left(  0,0,\sqrt{\Sigma\left(  1-\Sigma\right)  },\Sigma
,\sqrt{1-\Sigma},\ln\left(  \sqrt{\frac{\Sigma}{\gamma_{0}^{2}(1-\Sigma)}%
}\right)  \right)  ,\\
Q_{3}  &  =\left(  0,0,-\sqrt{\Sigma\left(  1-\Sigma\right)  },\Sigma
,\sqrt{1-\Sigma},\ln\left(  \sqrt{\frac{\Sigma}{\gamma_{0}^{2}(1-\Sigma)}%
}\right)  \right)  ,\\
Q_{4}  &  =\left(  0,0,0,1,0,\alpha\right)  ,\\
Q_{5}  &  =\left(  0,0,0,-1,0,\alpha\right)  ,
\end{align*}
which have the same physical properties as points $P_{1-5}$ respectively. The
de Sitter point, analogous to $P_{7}$, is recovered only when $\lambda=0$, and
it is the unique solution that describes cosmic acceleration.

\subsubsection{Stability properties}

Finally, as far as the stability is concerned, we recover the same conditions
as with the corresponding points $P_{1-5}$ and $P_{7}$.

\subsection{Analysis at Infinity}

While for the $SU\left(  2\right)  $ symmetry parameter
$\alpha$ is bounded, that is not true, for the case of $SL\left(
2,R\right)  $ symmetry. Therefore in order to investigate the
behaviour of the dynamical system when $\left\vert \alpha\right\vert
\rightarrow\infty$, we introduce the new variable $A=\tanh
\frac{\alpha}{2}$. Therefore, equation (\ref{ff.04}) becomes
\[
\frac{dA}{d\tau}=\frac{\sqrt{6}}{2}x\left(  1-A^{2}\right)  \text{.}%
\]
Therefore, when $A^{2}=1$, that is, at the infinity, there
are new stationary points with $x\neq0$. In this consideration, the
model $SL\left(  2,R\right)  $ model is reduced to a single scalar
field theory.

The new stationary points $R=R\left(  x,y,z,\Sigma,W,A\right)$ at the infinity are
\begin{align*}
R_{1}^{\pm}  & =\left(  -\frac{\lambda}{\sqrt{6}},\sqrt{1-\frac{\lambda^{2}%
}{6}},0,0,0,\pm1\right)  ,\\
R_{2}^{\pm}  & =\left(  -\sqrt{\frac{3}{2}}\frac{1}{\left\vert \lambda
\right\vert },\sqrt{\frac{3}{2}}\frac{1}{\left\vert \lambda\right\vert
},0,0,0\pm1\right)  ,\\
R_{3}  & =\left(  -\frac{2}{\sqrt{6}\lambda},\frac{2}{\sqrt{3}\lambda
},0,0,\sqrt{1-\frac{2}{\lambda^{2}}},-1\right)  .
\end{align*}
Stationary points $R_{1}^{\pm}$ describe a spatially flat
FLRW geometry with $q\left(  R_{1}^{\pm}\right)  =-1+\frac{\lambda^{2}}{2}%
$. The corresponding eigenvalues are $\left\{  -3+\frac{\lambda^{2}%
}{2},-3+\frac{\lambda^{2}}{2},\lambda^{2}-3,\frac{\lambda^{2}+2\lambda-6}%
{2},\frac{\lambda^{2}}{2}-1,\mp\lambda\right\}  $, which means that
$R_{1}^{+}$ is an attractor for $0<\lambda<\sqrt{2}$, while
$R_{+}^{-}$ is an attractor when $1-\sqrt{3}<\lambda<0$.

The stationary points $R_{2}^{\pm}$ describes FLRW solutions
with $q\left(  R_{2}^{\pm}\right)  =\frac{1}{2}$ and $\Omega
_{m}\left(  R_{2}^{\pm}\right)  =1-\frac{3}{\lambda^{2}}$. The
analysis of the eigenvalues reveal that the points are always saddle.

Finally, point $R_{3}$, corresponds to FLRW universe with
nonzero curvature, where the $q\left(  R_{3}\right)  =0$; 
the new point is an attractor for $\lambda>\sqrt{2}$.

\section{Conclusion}

\label{sec4}

In this work we investigated the cosmological dynamics of the nonlinear
$\sigma$-model in an anisotropic and homogeneous background geometry of the
Bianchi V type. We considered the two internal symmetries $SU(2)$ and
$SL(2,\mathbb{R})$, for which the kinetic metric of the multi-scalar field
configuration is a space of constant positive and negative curvature
respectively. The latter model is generally used to describe the so-called
hyperbolic inflation.

We employed an inhomogeneous ansatz for the profiles of the scalar fields.
Although the $\sigma$-model admits fewer symmetries than the base manifold,
the resulting equations are compatible with the Bianchi V background geometry.
As a potential function we introduced an exponential expression with respect
to one of the profile scalars, $V\left(  \alpha\right)  =V_{0}e^{\lambda
\alpha}$. Such potentials have played an important role in the description of
single-field and multi-field inflationary models. We additionally included a
dust fluid source with a four-velocity in the comoving frame.

The parameter $\gamma_{0}$, introduced through our ansatz to characterize the
inhomogeneity of the field, has been found to allow the existence of
anisotropic solutions in Bianchi V geometries. Indeed, in the homogeneous
limit $\gamma_{0}\rightarrow0$, the anisotropic parameter vanishes from the
nondiagonal component of the Einstein field equations, and the resulting
spacetime is either an open FLRW universe, or the spatial curvature vanishes
and the Bianchi I geometry is recovered. Furthermore, the parameter
$\gamma_{0}$ plays the role of a coupling parameter between the scalar fields
and the spatial curvature, while affecting the definition of the effective
mass term $m_{\alpha,eff}^{2}.$

For the asymptotic analysis we employed the Hubble normalization approach and
rewrote the field equations as an autonomous dynamical system, whose dimension
is reduced by the algebraic constraint (\ref{ff.07}). For the $SU(2)$ symmetry
we determined the stationary points $P_{1}-P_{6}$, which exist for arbitrary
values of the parameter $\lambda$. Point $P_{1}$ describes the isotropic
matter dominated solution and it is always a saddle point. Points $P_{2}$ and
$P_{3}$ describe anisotropic Bianchi V solutions with $q=2\Sigma$ and are
always unstable, while points $P_{4}$ and $P_{5}$ model unstable Bianchi I
solutions of the Kasner-like type with $q=2$. Point $P_{6}$ describes a
Bianchi V anisotropic cosmology where the nonlinear $\sigma$-model behaves as
the pressurless fluid source. Furthermore, on the limit $\lambda=0$, where the
potential reduces to the cosmological constant, an additional real stationary
point exists. It describes a de Sitter universe dominated by the cosmological
constant. This is the unique attractor of the system. 

Finally, the dynamical system analysis for the $SL(2,\mathbb{R})$ internal
symmetry of the nonlinear $\sigma$-model reveals similar conclusions to those
of the $SU\left(  2\right)  $ model. However, in the $SL(2,\mathbb{R})$ model,
there exist a new class of solutions where the model is reduced to a single
scalar field model which allow the existence of slow-roll inflationary
solutions. 

By comparing these results with those of the homogeneous case, it is
clear that inhomogeneities prevent the existence of the two possible
accelerated phases which suppress the cosmological anisotropies provided by
the nonlinear $\sigma$-model. Hyperbolic inflation is not recovered
\cite{ans1}, while slow-roll inflation is recovered only for the 
$SL(2,R)$ internal symmetry in the extreme limits where 
$\phi\rightarrow\pm\infty$.

In a future work, we plan to investigate whether this conclusion holds
true also for other Bianchi models of Class B. Finally, the effects of
inhomogeneities in conjunction with a titled fluid will be examined elsewhere.

\begin{acknowledgments}
AG was supported by Proyecto Fondecyt Regular 1240247. GL \& AP acknowledges
the support from Proyecto Fondecyt Regular 1240514. GL thanks the support of
Vicerrectoria de Investigacion y Desarrollo Tecnologico (VRIDT) of Universidad
Catolica del Norte (UCN) through Resoluci\'{o}n VRIDT No. 096/2022,
Resoluci\'{o}n VRIDT No. 200/2025, and Resoluci\'{o}n VRIDT No. 021/2026. The
authors acknowledge the COST Action CA23130 \textquotedblleft Bridging high
and low energies in search of quantum gravity (BridgeQG)\textquotedblright.
\end{acknowledgments}

\end{document}